\documentclass{jaa}
\usepackage{natbib}
\usepackage{deluxetable}
\usepackage{graphicx}	
\usepackage{amsmath}	
\usepackage{amssymb}
\usepackage{times}
\usepackage{latexsym}
\usepackage{fnpos}
\usepackage{scrextend}
\usepackage{scalerel}
\usepackage{hyperref}
\usepackage{tabularx}
\usepackage{xspace}
\usepackage{enumerate}
\usepackage{longtable}
\usepackage{lscape}

\def\aj{Astron. J.}

\def\apj{Astrophys. J.}
\def\apjl{Astrophys. J. Lett.}
\def\apjs{Astrophys. J. Suppl.}
\def\aap{Astron. Astrophys.}
\def\aaps{Astron. Astrophys. Suppl.}
\def\mnras{Mon. Not. R. Astron. Soc.}
\def\nat{Nature}
\def\pasa{Pub. Astron. Soc. Aus.}

\def\aaps{A\&AS} 

\begin{document}\sloppy

\title{The central region of the enigmatic Malin 1}


\author{Kanak Saha\textsuperscript{1,*}, Suraj Dhiwar\textsuperscript{2,3}, Sudhanshu Barway\textsuperscript{4}, Chaitra Narayan\textsuperscript{5}, Shyam N. Tandon\textsuperscript{1}}
\affilOne{\textsuperscript{1}Inter-University Centre for Astronomy and Astrophysics, Pune 411007, India.\\}
\affilTwo{\textsuperscript{2}Department of Physics, Savitribai Phule Pune University, Pune 411007, India\\}
\affilThree{\textsuperscript{3}Dayanand Science College, Barshi Road, Latur, Maharashtra 413512, India\\}
\affilFour{\textsuperscript{4}Indian Institute of Astrophysics (IIA), II Block, Koramangala, Bengaluru 560 034, India.\\}
\affilFive{\textsuperscript{5}National Centre for Radio Astrophysics-TIFR, Pune, India.\\}

\twocolumn[{

\maketitle

\corres{kanak@iucaa.in}

\msinfo{10 10 2020}{10 10 2020}

\begin{abstract}
Malin 1, being a class of giant low surface galaxies, continues to surprise us even today. The HST/F814W observation has shown that the central region of Malin 1 is more like a normal SB0/a galaxy, while the rest of the disk has the characteristic of a low surface brightness system. The AstroSat/UVIT observations suggest scattered recent star formation activity all over the disk, especially along the spiral arms. The central 9" ($\sim 14$~kpc) region, similar to the size of the Milky Way's stellar disk, has a number of far-UV clumps - indicating recent star-formation activity. The high resolution UVIT/F154W image reveals far-UV emission within the bar region ($\sim 4$~kpc) - suggesting the presence of hot, young stars in the bar. These young stars from the bar region are perhaps responsible for producing the strong emission lines such as H$\alpha$, [O{\sc{ii}}] seen in the SDSS spectra. Malin 1B, a dwarf early-type galaxy, is interacting with the central region and probably responsible for inducing the recent star-formation activity in this galaxy.

\end{abstract}

\keywords{glaxies:individual: Malin 1 ---galaxies: structure---galaxies:evolution---galaxies:interaction ---galaxies:formation}

}]


\doinum{12.3456/s78910-011-012-3}
\artcitid{\#\#\#\#}
\volnum{000}
\year{0000}
\pgrange{1--}
\setcounter{page}{1}
\lp{1}

\section{Introduction}
\label{sec:introduction}
Since it's discovery \citep{Bothunetal1987}, Malin~1 continues to surprise us. Malin~1 belongs to the class of giant low-surface brightness (GLSB) galaxies and probably, is one of the largest disk galaxies with spiral arms extending upto 130 kpc in radius at the distance of Malin~1 \citep{MooreParker06, Boissier2016}. Most of the stars in the outer disk of Malin~1 follows an exponentially declining surface brightness profile. The outer disk is so faint, that when extrapolated to the centre, it has a central surface brightness of 25.5 mag~arcsec$^{-2}$ in the V-band \citep{ImpeyBothun1989}. Using the deep images in the six photometric bands from the NGVS and GUViCS surveys \cite{Boissier2016} have shown that the extended low-surface brightness disk has a long and low star-formation history. However, the central region has a complex structure - with a bulge and a bar as revealed by the Hubble Space Telescope (HST) observation \citep{Barth2007}. In other words, the morphology of the central region is similar to the high surface brightness (HSB) galaxies with SB0/a morphology. This, in turn, makes Malin~1 a classic example of a hybrid galaxy with central HSB-like and outer LSB-like envelope. The central HSB-like structure is also manifested in terms of the inner steeply rising rotation curve \citep{Lellietal2010}. Several questions arise here -- is Malin~1 a galaxy in transition from LSB to HSB? What drives such a transformation? In addition to stars, the galaxy contains plenty of neutral hydrogen gas \citep{Pickeringetal1997} as in typical late-type spirals. Even then, it appears that the galaxy may not be forming stars efficiently as it lacks CO molecules \citep{Braineetal2000}, although, the non-detection of CO may also be attributed to the low temperature of molecular gas compounded with sub-solar metallicity \citep{Braineetal2000}. Nevertheless, the central region appears to have recent star formation activity - based on the strong emission lines such as H$\alpha$, H$\beta$, [O{\sc ii}], [O{\sc iii}] from the SDSS central 3" fibre spectra and much recently from a detailed spectroscopic study by \cite{Junaisetal2020}. Such an activity is probably induced by an external perturber.

Indeed, Malin~1 is interacting with another galaxy called Malin~1B which an early type galaxy at a distance of only 9" (14 kpc) from the centre. Recently \cite{Reshetnikovetal2010} suggest that Malin~1 is also probably interacting with another galaxy SDSSJ123708.91+142253.2 at a distance of about 358 kpc away. Such an interaction can lead to the formation of the central bar \citep{MiwaNoguchi1998,Langetal2014,Lokasetal2014}. Accretion of smaller galaxies like Malin~1B or giant clumps could enhance star-formation, increase the central concentration - in other words, grow the central bulge \citep{Aguerrietal2001,Eliche-Moraletal2018}. One might expect to see the induced star-formation activity in the UV emission and indeed, this is evident from the far-UV (FUV) and near-UV (NUV) observation of this galaxy by GALEX. However, due to the large 5" PSF, the central 9" region remains practically unresolved in both FUV and NUV filters.
To achieve a detailed understanding of the recent star-formation activity and the nature of the central disk, we observed Malin~1 with the Ultra-Violet Imaging Telescope (UVIT) on-board AstroSat, simultaneously in far-ultraviolet (FUV) and near ultraviolet (NUV) bands. Our analyses show that there is clumpy star-formation in the central disk hosting a bar and S0 morphology. We also find evidence of prominent emission in the FUV band from the bar region of the galaxy. 

The paper is organized as follows - Sec~\ref{sec:obs} describes UVIT observation and data reduction. The treatment we use to correct for dust extinction is briefly described in Sec~\ref{sec:dust}. In Sec~\ref{sec:comp}, we compare the UVIT observations with that of GALEX and CFHT. We discuss the color composition of the central bulge and bar region in Sec~\ref{sec:morphology} and a detailed decomposition of the optical surface brightness profile in Sec~\ref{sec:decomposition}. In Sec~\ref{sec:UVandHI}, we cover the UV surface brightness as well as the HI emission. We discuss in detail on the central bar and the central star formation activity in Sec~\ref{sec:bar}. Finally, in Sec~\ref{sec:conclusion} we summarise and conclude. 

\section{UVIT observation and data analysis}
\label{sec:obs}

The giant low-surface brightness galaxy Malin 1 (at a redshift of z=0.0827) was observed simultaneously in the far-UV (F154W; 1250 - 1750\AA) and near-UV (N263M; 2461 -2845\AA) filters by the UVIT during February 2017 (PI: Kanak Saha; proposal id: G06-016). The observations were carried out in the photon counting mode (frames taken every 34 millisecond during each orbit) resulting in about $\sim 45000$ frames accumulation in a typical good dump orbit. The orbit-wise L1 dataset was processed using the official L2 pipeline. The pipeline throws away the cosmic-ray affected frames during data reduction process and these frames were excluded in the final science-ready images in FUV and NUV bands and the subsequent calculation of the photometry. The final science-ready images had a total exposure time of $t_{exp}=9850$~sec in F154W (bandwidth=380\AA) and $t_{exp}=9600$~sec in N263M (bandwidth=275\AA).  

\par
Astrometric correction was performed using the {\em GALEX} and {\em CFHT} u band images as the reference images. We have used an IDL program which takes an input set of matched xpixel/ ypixel from {{\em UVIT}} images (F154W and N263M) and RA/ Dec from {\em GALEX} and {\em CFHT} u band images. We then perform a TANGENT-Plane astrometric plate solution similar to ccmap task of IRAF \citep{Tody93}. The astrometric accuracy in NUV was found to be $\sim 0.15"$ while for FUV, the RMS was found to be $\sim 0.22"$, approximately half a (sub)pixel size. 
Note that the absolute astrometric accuracy for GALEX based on QSOs data is about 0.5" \citep{Morrisseyetal2007}. It is also known that there is systematic offset between SDSS and GALEX $\sim$ 0.1 - 0.3".

\begin{figure*}
  \includegraphics[height=.65\textheight]{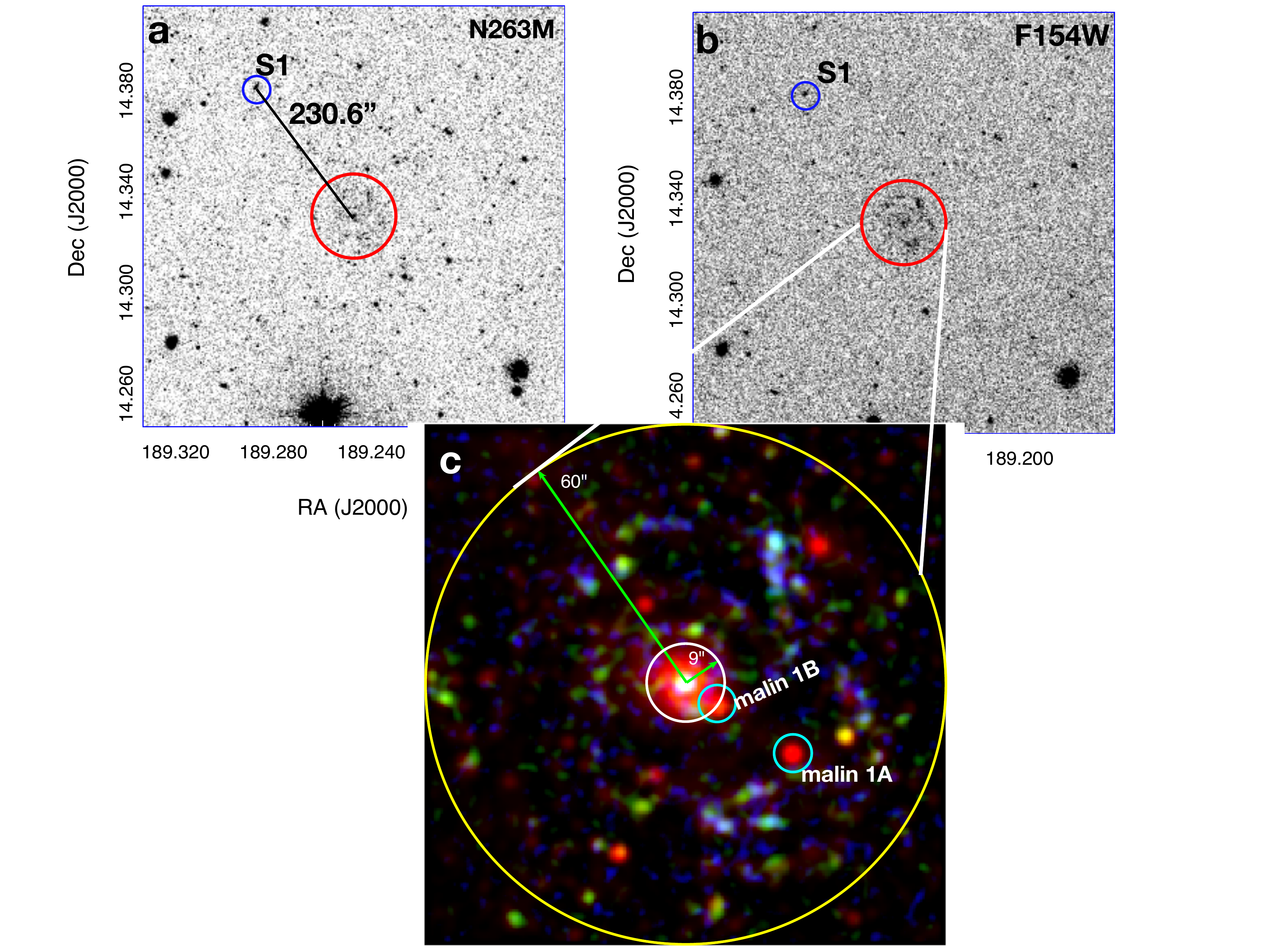}
  \caption{Top panels (a,b): AstroSat/UVIT observation of Malin 1. S1 (at a distance of 230.6"=358 kpc) is believed to be interacting with Malin 1. Bottom panel: An RGB colour image of the region marked by red circle - using CHFT g-band(red), AstroSat N263M (green) and F154W (blue). Both N263M and CFHT g-band images are convolved with F154W PSF and resampled to F154W pixel scale.}  
\label{fig:uvitimage}
\end{figure*}

The photometric calibration is performed with a white dwarf star Hz4; the photometric zero-points are 17.78 and 18.18 for F154W and N263M respectively \citep{Tandonetal2017a}. Once photometric calibration and astrometric correction are successfully applied, we extract an image of size 600" x 600" as shown in Fig.\ref{fig:uvitimage}. The morphology and color of the galaxy are discussed in sec.~\ref{sec:morphology} We run SExtractor~\citep{BertinArnouts1996} on this cutout image of size 600" and extract the $3\sigma$ sources. After removing the $3\sigma$ sources, we place a large number of random apertures of $7\times7$ pixel boxes avoiding the location of the extracted sources. This was repeated with boxes of size 11 pixels. We use the mean of the resulting histogram (nearly symmetric) as the background. For F154W, the background is $B_{f}=9.5 \times 10^{-6}$~ct~s$^{-1}$~pix$^{-1}$ with a $\sigma_{f}=3.94 \times 10^{-5}$~ct~s$^{-1}$~pix$^{-1}$. In N263M filter, the sky background is $B_{n}=7.24 \times 10^{-5}$~ct~s$^{-1}$~pix$^{-1}$ with a $\sigma_{n}=6.48 \times 10^{-5}$~ct~s$^{-1}$~pix$^{-1}$.
Note that the sextractor background from the FUV image is even lower and it is $B=1.15 \times 10^{-8}$~ct~s$^{-1}$~pix$^{-1}$ with an rms of $1.9 \times 10^{-5}$~ct~s$^{-1}$~pix$^{-1}$. We did not consider this in our subsequent calculation.

\subsection{PSF Modeling}
\begin{figure}
\includegraphics[height=.42\textheight]{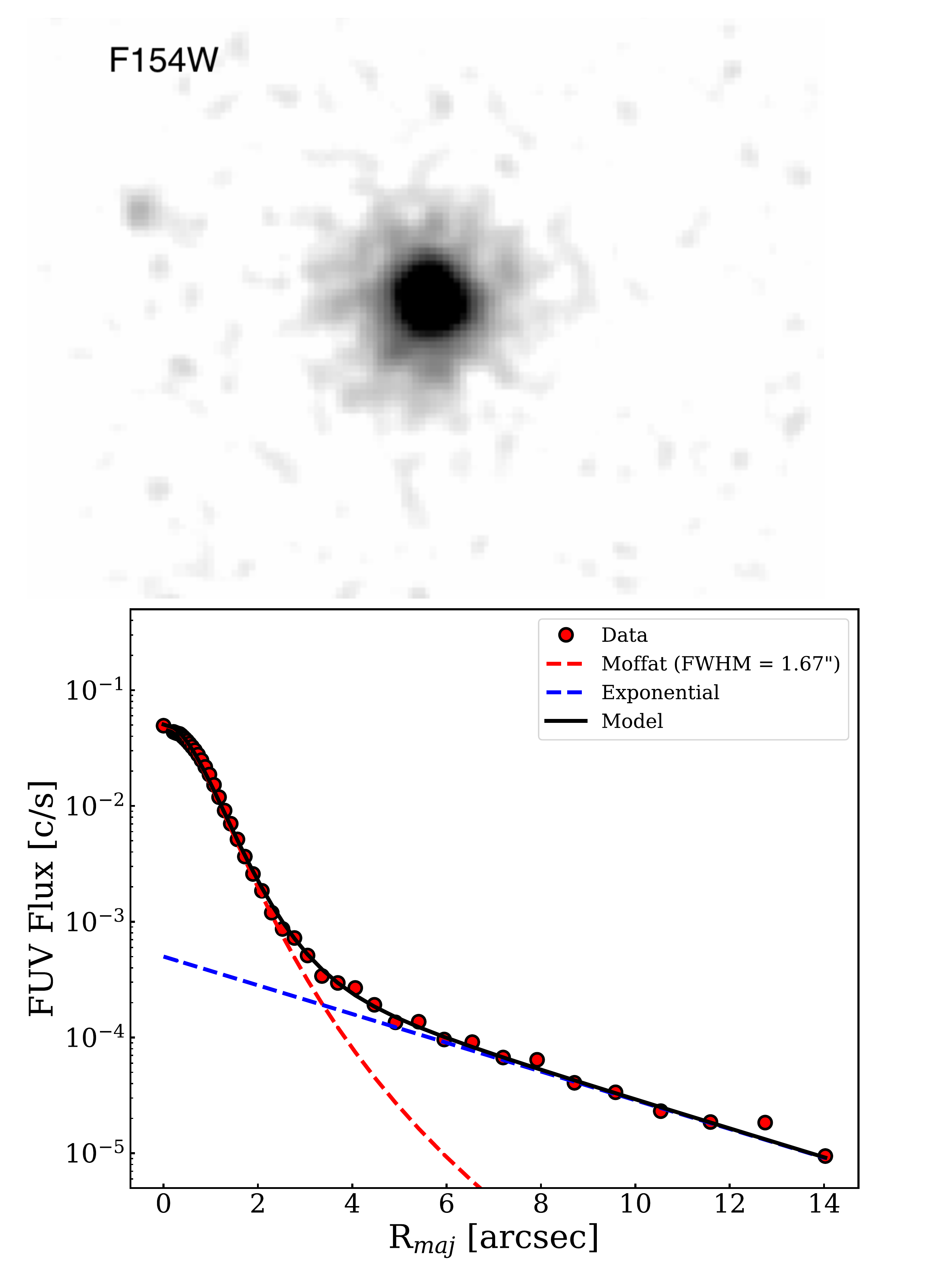}
\caption{Upper panel: image of the star in F154W filter. Bottom panel: PSF modeled with a moffat and an exponential profile for the wing.}
\label{fig:psf}
\end{figure}

FUV modelled PSF is shown in Figure~\ref{fig:psf}. To model the PSF and estimate the FWHM, we have chosen a bright star (RA: 189.2595, Dec:14.2518) right at the bottom of Malin 1 (see Fig.~\ref{fig:uvitimage}). We used IRAF \citep{Tody93} to obtain the surface brightness profiles (SBP) of the star in FUV band. The star in NUV is heavily saturated, so we have avoided a discussion of the NUV PSF here. Instead, we have considered a circular symmetric PSF from the recent calibration paper \citep{Tandonetal2020} for NUV modelling. The SBP reveals long wing outside the core in FUV (see the star's image in the upper panels of Fig.\ref{fig:psf}). The FUV star profile is fitted with two components: a Moffat profile for the core, and an exponentially declining profile for the wing as follows:
\par

\begin{equation}
     I_{PSF}(r) = I_{m0} \left[1 + \bigg(\frac{ r } { \alpha }\bigg)^2  \right]^{-\beta} 
 + I_{w0} e^{-\frac{r}{h_{0}}}, 
\label{eq:psf}
\end{equation}

\noindent where $I_{m0}$ and $I_{w0}$ denote the central intensity values of the inner Moffat and outer exponential profiles respectively. For the Moffat, the FWHM = $ 2\alpha \sqrt{2^\frac{1}{\beta}-1}$; where $\beta$ and $\alpha$ are the free parameters, $\beta$ determining the spread of the Moffat function. Note the Gaussian function is a limiting case of Moffat function (${\beta} \rightarrow {\infty}$).

 The best fit model was obtained with two components (Moffat and exponential). With the moffat fit, the estimated PSF obtained is 1.67" with $\beta$ = 1.4. We discuss the wing of the PSF separately below.

\subsection{PSF wing}
The wing of the PSF in F154W follows an exponential fall-off. This is also true for the N263M band where the exponential profile extends upto $\sim 35"$ or $\sim 84$~pixels (not shown here). In F154W, the wing has been modelled till $\sim 14"$ or $\sim 34$~pixels.  
In F154W, the scale-length of the exponential profile is $h_{0}=3.5"$ or $8.4$~pixels.  Accurate estimation of the faint emission from the galaxy will require a thorough knowledge of the PSF wing which reflects amount of the scattered light in the same wavelength. We found that the wing of the PSF contains about $19\%$ of the total flux while 81\% is in the core in accordance with recent UVIT calibration \citep{Tandonetal2020}.

\subsection{Dust correction}
\label{sec:dust}

Both FUV and NUV fluxes are susceptible to both foreground and internal dust extinction. \citet{Schlegel98} full-sky 100 $\mu$m map provides us with the foreground dust reddening $E(B - V)$ along a given line of sight. The mean $E(B - V) =0.038$ along the direction of Malin 1. The values of foreground extinction parameter for the FUV and NUV bands are then derived using $A_{\lambda} = k_{\lambda} E(B-V)$; where $k_{\lambda}$ is obtained following Calzetti extinction law \citep{Calzettietal2000}. For F154W, $A_{F154W}=0.39$~mag with $k_{F154W}=10.18$ while for N263M, $A_{N263M}=0.29$~mag with $k_{N263M}=7.57$.   

The nebular colour excess can be estimated using the method of Balmer decrement \citep{OsterbrockFerland2006} assuming case B recombination, temperature $T=10^4$~K and electron density $n_e = 100 cm^{-3}$ as: 

\begin{equation} 
E(B-V) = 1.97 \log\big{[}\frac{(H{\alpha}/H{\beta})_{obs}}{2.86}\big{]},
\label{eq:balmer}
\end{equation}

\noindent where H$\alpha$ and H$\beta$ are the observed line fluxes. Based on the SDSS spectra of the central 3" region, the observed H$\alpha$ and H$\beta$ fluxes are $224.5\pm 5.8 \times 10^{-17}$ and $67\pm 4.6 \times 10^{-17}$~erg~s$^{-1}$~cm$^{-2}$ respectively. The internal color excess $E(B-V)_{nebular}=0.1323$. In the subsequent calculation, we use the Calzetti relation \citep{Calzettietal2000} for stellar continuum color excess $E(B-V)_{star}=0.44\times E(B-V)_{nebular}=0.058$ - suggesting low extinction in the central region of the galaxy \citep{Junaisetal2020}. This is in-sync with general trend found in LSB galaxies which are often devoid of dust and molecular gas compared to their HSB counterparts \citep{Rahmanetal2007, Dasetal2006}. In the rest of the calculation, we use these values of extinction for the full galaxy.

\begin{figure*}
\begin{flushleft}
\includegraphics[height=.77\textheight]{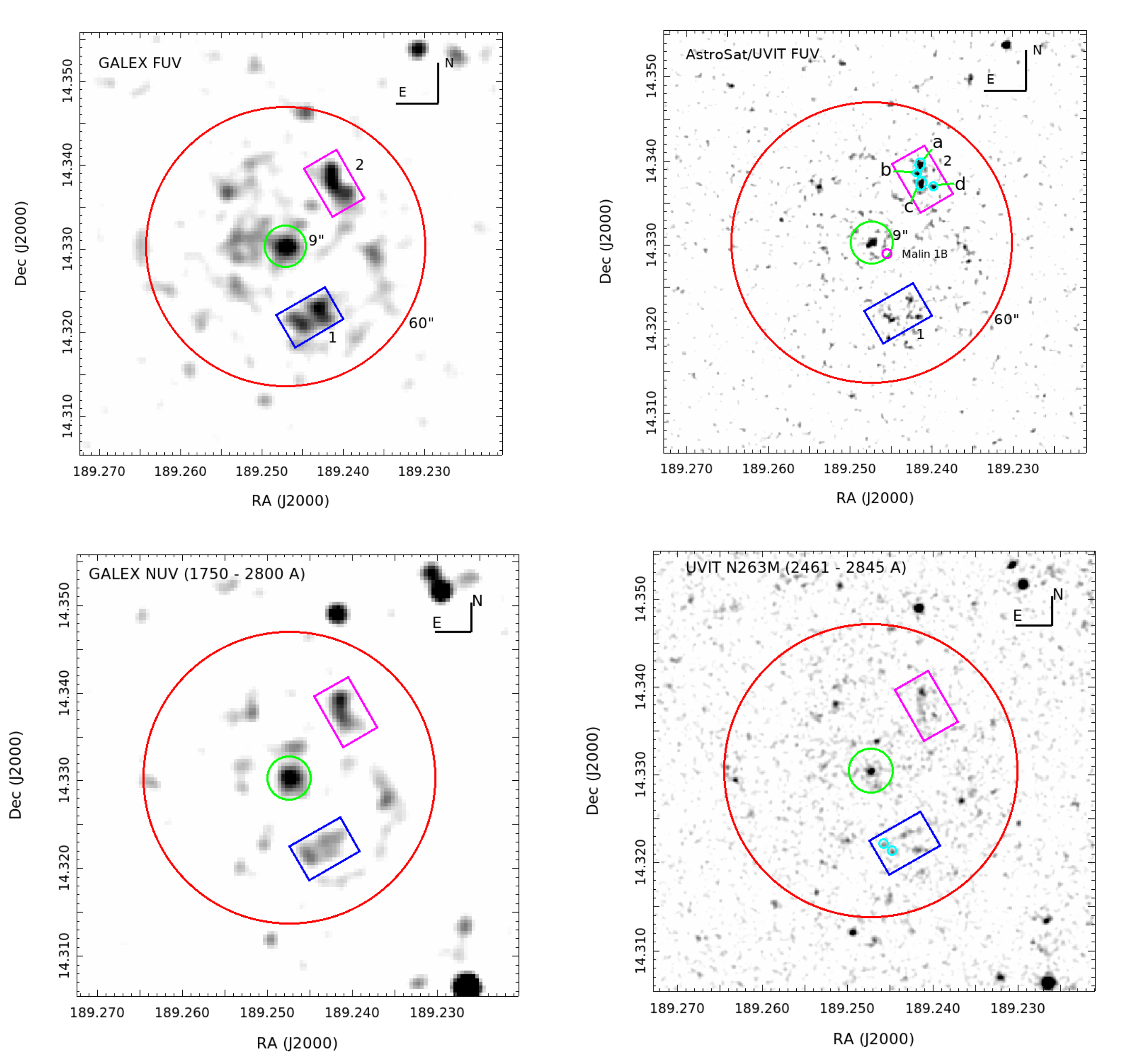}
\caption{A comparison between the GALEX FUV and AstroSat/UVIT FUV image (upper panels). The bottom panels show the NUV images. The radius of the red circle centred on Malin 1 is 60" and that of the inner green one is 9". On the UVIT FUV image (top right panel), Malin 1B (RA: 12:36:58.83, Dec: +14:19:44.11) is marked by the magenta circle (2" radius). In the region 2 (top right panel), we have marked four clumps a, b,c and d. Boxes are of size 24" x 16 ".}
\label{fig:GALEX-UVIT}
\end{flushleft}
\end{figure*}

\begin{figure*}
\includegraphics[height=.35\textheight]{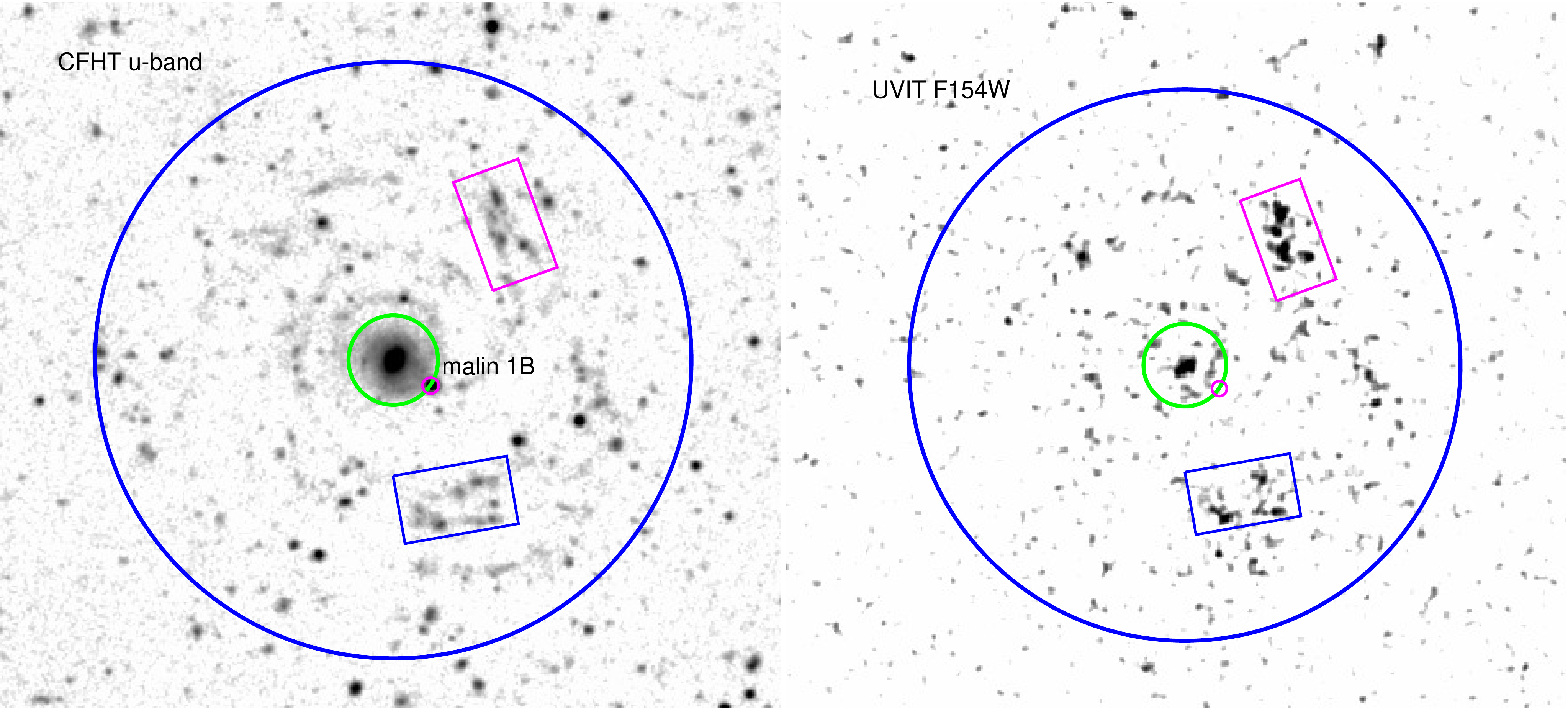}
\caption{A comparison between the CFHT u-band and AstroSat/UVIT FUV image. The radius of the blue circle centred on Malin 1 is 60" and that of the inner green one is 9". The magenta circle marks the location of Malin 1B.}
\label{fig:cfht-UVIT}
\end{figure*}

\subsection{Comparison with GALEX observation}
\label{sec:comp}
The FUV image from GALEX has an exposure time of 3154 sec while in  UVIT/F154W band, it is 9850 sec. Compared to GALEX NUV, our NUV band N263M is narrower and avoids the 2175\AA bump in the extinction curve. In Fig.~\ref{fig:GALEX-UVIT}, we show a one-to-one comparison between the GALEX and UVIT observation of Malin 1. Rectangular regions marked 1 and 2 along the spiral arms of Malin 1 in the GALEX images are resolved into clumps in the UVIT images. With the UVIT FUV image, it is thus possible to study the young star-forming clumps along the spiral arms. The region 2 in GALEX FUV image resolves into four clear clumps (marked by a, b, c, and d) in UVIT F154W image. We have measured their aperture magnitudes and the corresponding signal-to-noise ratios. The faintest of the four far-UV clumps {\bf b}, has a background subtracted magnitude of 24.27 AB mag with a S/N=4.2. While the brightest among them is the clump {\bf c} having a magnitude of 23.3 AB mag and S/N=7.3. 
One of the clumps marked by cyan circle in region 1 (bottom right panel of Fig.~\ref{fig:GALEX-UVIT}) in the N263M band has been detected at S/N=7.2 and has a magnitude of 23.65 AB mag within a circular aperture of radius 1.6".

Due to the improved image resolution and sensitivity of UVIT, we are able to measure the far-UV flux from Malin 1B , marked by magenta circle, located at a distance of 9" from the centre (see Fig.~\ref{fig:GALEX-UVIT} and ~\ref{fig:cfht-UVIT}). Within an aperture of radius 2", Malin 1B has a magnitude of 24.9 AB mag (and 24.5 AB mag after foreground correction) with S/N=3.3. In comparison, Malin 1B is completely blended in the GALEX FUV as well as in the NUV images (see the left panels of Fig.~\ref{fig:GALEX-UVIT}). The inner regions (denoted by green circle with radius $=9"$) in the GALEX images are also not resolved. Whereas in their UVIT counterpart, it is possible to locate a few clumps in the FUV F154W band and N263M band - indicating recent star-forming activity. A closer inspection of the central region of the F154W image shows that there is elongated structure in FUV; when zoomed into this, we found two clumps aligned to give an impression of a bar-like structure. We discuss this structure in detail in the following section. Interestingly, the bar-like structure is also seen in the N263M (relatively dust-free); although it appears like a elliptical blob.

\begin{figure}
\begin{flushleft}
  \includegraphics[height=0.33\textheight]{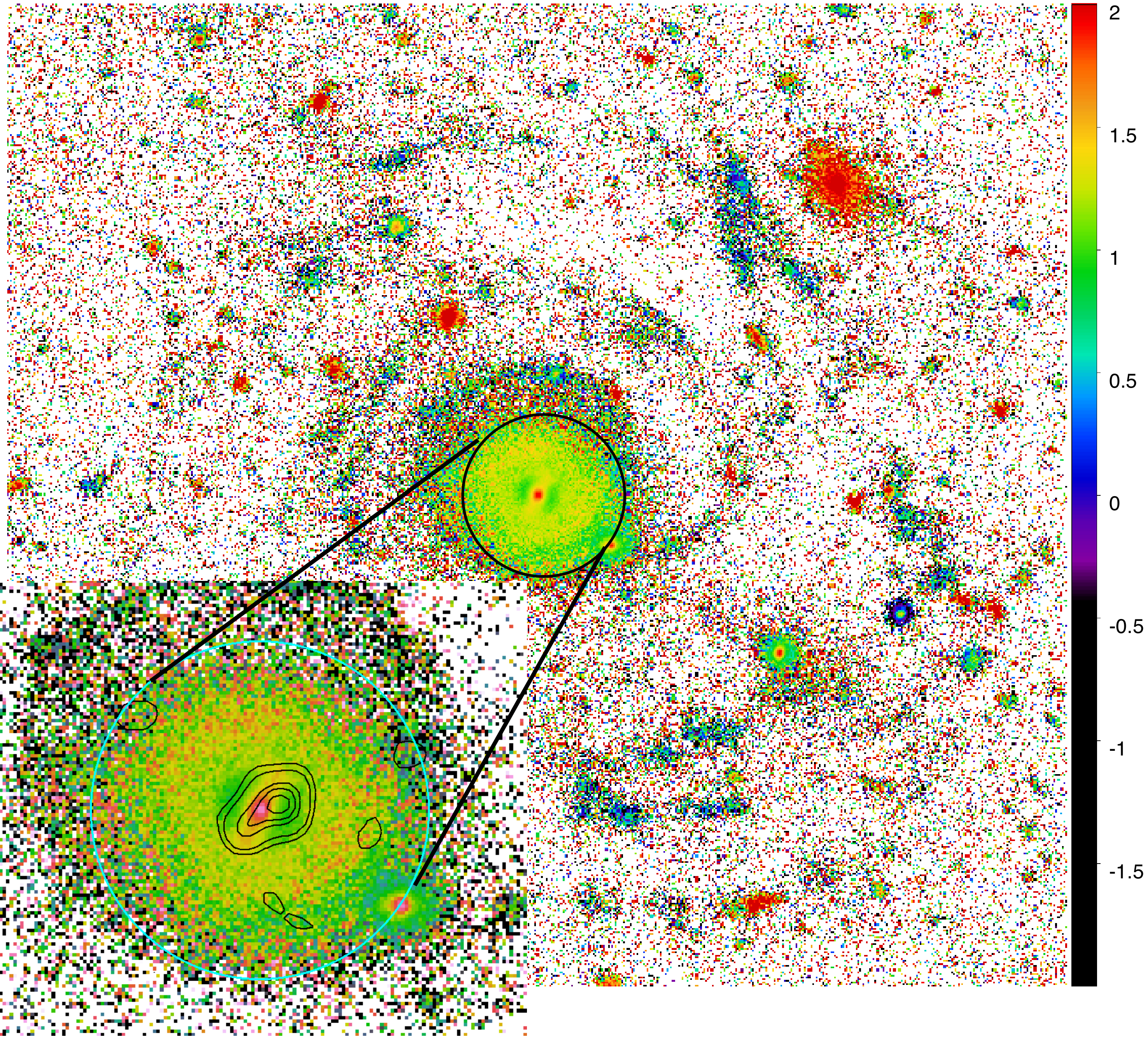}
  \caption{CFHT g-i color map of Malin 1 and its central 9" region. The unit of the colorbar is in magnitude. Inset figure highlights the central disk with F154W contour of the bar region overlaid on the g-i color map. A part of the FUV emission appears misaligned with respect to the bar.}
  \label{fig:colormap}
\end{flushleft}
\end{figure}

In Fig.~\ref{fig:cfht-UVIT}, we compare the UVIT/FUV morphology with the CFHT u-band image. The better resolution and depth of the CFHT observation \citep{Boissier2016} makes it possible to compare both regions 1 and 2 in considerable detail. The star-forming far-UV clumps (a,b,c,d) marked in Fig.~\ref{fig:GALEX-UVIT} are clearly visible in the CFHT u, g and i bands. The observation of these clumps in the optical bands, in particular the i-band, alongside the far-UV, indicates that they are older than 100 Myr. In other words, the spiral arms in Malin 1 although appear bluer, are not as young as a few 100 Myr. The combination of optical and far-UV emission suggests that there is a mixture of both young and old stellar populations along the spirals arms e.g., see regions 1 and 2 as well as the color-map in Fig.~\ref{fig:colormap}.   

\section{Morphology and colormap}
\label{sec:morphology}
The bottom panel of Fig.~\ref{fig:uvitimage} shows a color composite of Malin 1 in CFHT g-band and the two UVIT filters. The spiral arms extend to large radii - it is possible to trace it clearly up to a radius of 60"=93 kpc, perhaps even larger radii \citet{Boissier2016}. Interestingly, the spiral arms are not wide open, like in NGC 1566 \citep{Gouliermisetal2017}. The CFHT g-band image reveals more of a tightly wound type, although it is believed to be interacting with an external galaxy SDSS J123708.91+142253.2 \citep{Reshetnikovetal2010} located at a distance of about 358 kpc away. Apart from the spiral arms, the disk in the outer part is nearly invisible - in other words, even with the deep optical images from CFHT and Megacam on Megallan telescope \citep{Ferrareseetal2012,Galazetal2015,Boissier2016}, we just see the spiral arms with some diffuse stellar streams down to 28 mag~arcsec$^{-2}$ in the B-band \citep{Galazetal2015}. These spiral arms are asymmetric \citep{MooreParker06}, both in radii and azimuths; the asymmetry is prominent along the North-East and South-West direction (Fig.~\ref{fig:uvitimage}). One can easily trace the F154W light (in blue) along the same direction. However, as one moves inward, the galaxy morphology changes, rather dramatically. There is clearly a brighter component in the central $9" = 14$~kpc region. Interestingly, this central region is nearly the same size as that of the stellar disk of the Milky Way. As revealed in the HST observation, this central region harbours a bar \citep{Barth2007}. In Fig.~\ref{fig:colormap}, we show the 2D g-i color map of the galaxy using the CFHT deep observation and the FUV-NUV color from AstroSat/UVIT. Globally, the galaxy has a clear color gradient with bluer outskirts, see \cite{Boissier2016}. By comparing with Fig.~\ref{fig:GALEX-UVIT}, one can easily trace the regions 1 and 2 with bluer g-i color. However, the central 9" region is comparatively redder with $g-i \sim 1 - 1.5$, except the central bulge. The colormap reveals the central bar very clearly with $g-i \simeq 1.29$. The central bulge is red with $g-i \sim 2$- suggesting old stellar population. The comparative bluer g-i color of the bar in Malin 1 suggest triggering of star formation in the bar region given the Malin 1 seems to be interacting with the neighbour Malin 1B \citep{BarwaySaha2020}. The evidence of star-formation activity in the bar region is also clear from the FUV contours shown on the optical g-i color map. The $FUV-NUV$ color map of Malin~1 is noisy. There are several pixels with zero count/s in the FUV observation. On a visual inspection, it appears that the overall $FUV-NUV \simeq 0$, although the very central part is redder. In the subsequent analysis, we do not use this $FUV-NUV$ colormap to draw any inference. 

\begin{figure*}
\includegraphics[height=.37\textheight]{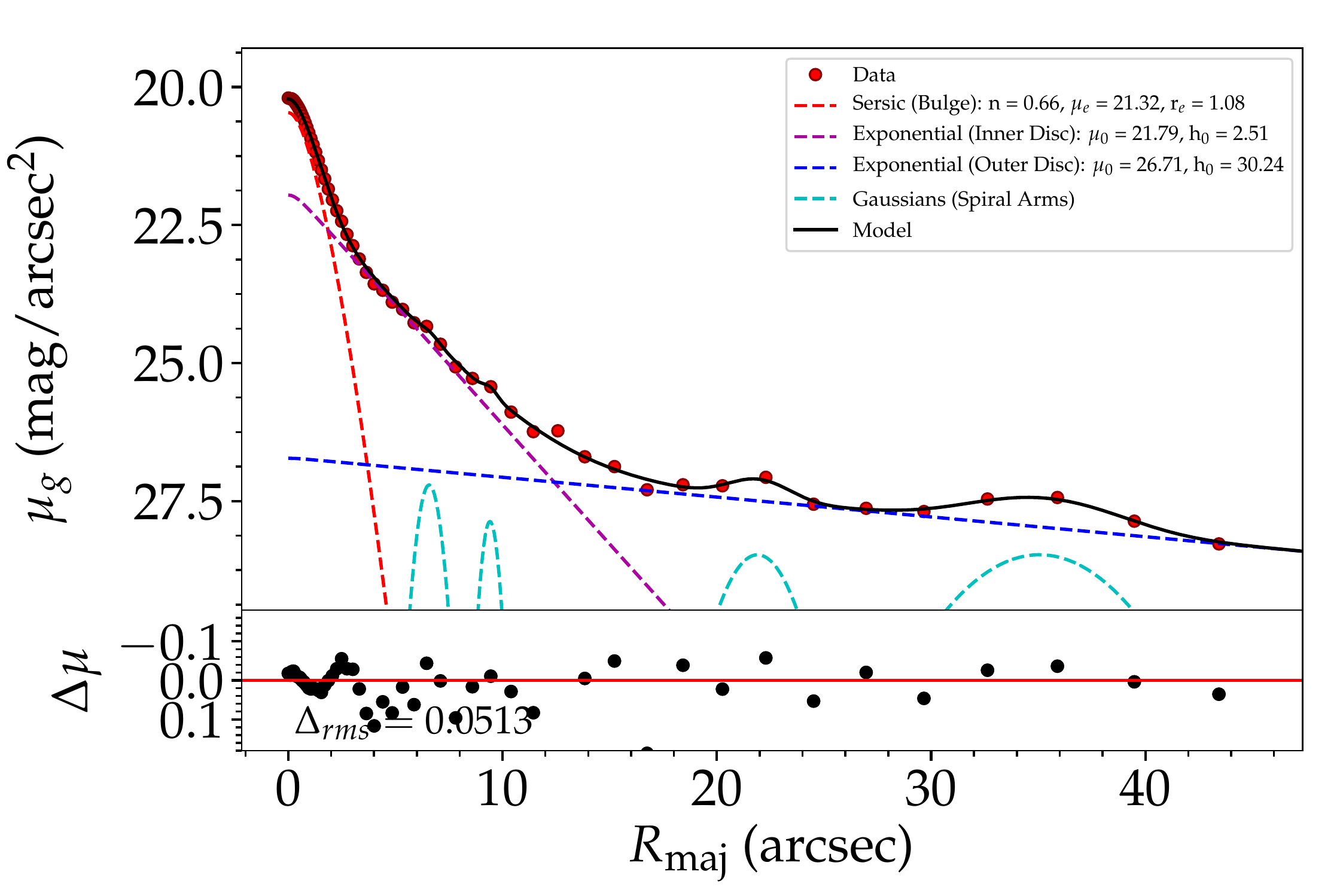}
\caption{CFHT g band surface brightness profile fitted with a sersic function for the bulge (dashed red line), and two exponential components (dashed purple and blue lines) and spiral arms and non-uniformities fitted by Gaussians (dashed cyan lines). The black solid line shows the best fit model.}
\label{fig:fullSBP}
\end{figure*}

\section{Multi-component decomposition}
\label{sec:decomposition}

In the following, we obtain the surface brightness profile (SBP) using the IRAF ellipse task on the CFHT g-band image (Fig.~\ref{fig:fullSBP}). We model this SBP using a combination of the standard Sersic profile \citep{Sersic1968} for the bulge and exponential profile \citep{Freeman1970} for the stellar disk. We also use the Sersic profile to model the central bar as in \cite{PahwaSaha2018}. The Sersic + exponential profiles are convolved with a Gaussian PSF of FWHM~0.7" (in CFHT g-band) and the resulting model is fitted to the observed g-band SBP using profiler \citep{Ciambur16}. 

\subsection{Faint exponential envelope}

The 1D profile decomposition (Fig.~\ref{fig:fullSBP}) shows that there is a very faint low-surface brightness exponential disk whose central surface brightness is $\mu_{0,g}=26.7$~mag~arcsec$^{-2}$ and a large scale length $h_{0}=30.2" \sim 47$~kpc. The origin of such a faint and diffuse stellar disk is elusive. Could it be due to the interaction with the SDSS galaxy \citep{Reshetnikovetal2010} or due to the accretion of Malin 1B, a dwarf early-type galaxy which in-spiraled towards the central region as a result of the dynamical friction? The outer region beyond 20" has spiral arms as can be seen by two kinks on the radial light profile - one at $\sim 22"$ and another at $\sim 35"$. From the nature of the spiral arms (more or less like a tightly wound), it might be possible to infer that Malin 1B might have excited it during its inward in-spiraling movement. Interestingly, the spiral arms do not extend all the way to the central region; it stops around $8- 10"$ from the center and this is apparently the high surface brightness region of the galaxy with a morphology of SB0/a. It might be possible that this hotter inner region is acting like a Q-barrier for sustaining the spiral structure in the outer disk \citep{SahaElmegreen2016}.

\subsection{Intermediate exponential disk}

The slope of the SBP changes significantly at radii below 20". We fit the intermediate region (from about 6 - 20") with an exponential profile motivated by the nature of the profile in this region. The central surface brightness of this disk is $21.79$~mag~arcsec$^{-2}$ in g-band and has a scale length of 2.51"=3.9~kpc. Considering the B-band $\mu_{0}$=22.5~mag~arcsec$^{-2}$ and the same in r-band as 21~mag~arcsec$^{-2}$ as the definition of LSB disk \citep{PahwaSaha2018}, we obtain $\mu_{0}=21.9$~mag~arcsec$^{-2}$ in the g-band for the definition of LSB disk. According to this and the comparatively larger scale-length, the intermediate disk is also an LSB disk or a marginal LSB disk.
\par
The best-fit parameter reveals a compact pseudo-bulge at the central region with an effective radius $r_{e}=1.08"=1.6$~kpc and Sersic index $n=0.66$. However, the central region is better modelled with the higher resolution images from HST as Malin 1 has been observed in the HST/WFPC2/F300W and F814W bands, see \cite{Barth2007}. Before we discuss the central region and the bar in detail, it is useful to understand the global UV surface brightness profile (especially the central region) and cold neutral hydrogen gas distribution in the galaxy.

\section{UV surface brightness profile and H{\sc i}}
\label{sec:UVandHI}
\cite{Boissier2016} have performed a detailed modelling of the FUV and NUV surface brightness profiles using GALEX observation. Although they exclude the central 20" region from their analysis, measurements of FUV and NUV surface brightness levels are shown at 10" and they are about $27$~mag~arcsec$^{-2}$. Whereas at a radius of 20" from the centre the FUV and NUV surface brightness levels are $28.7$ and $28.5$~mag~arcsec$^{-2}$ respectively.

\begin{figure}
\begin{flushleft}
  \includegraphics[height=.32\textheight]{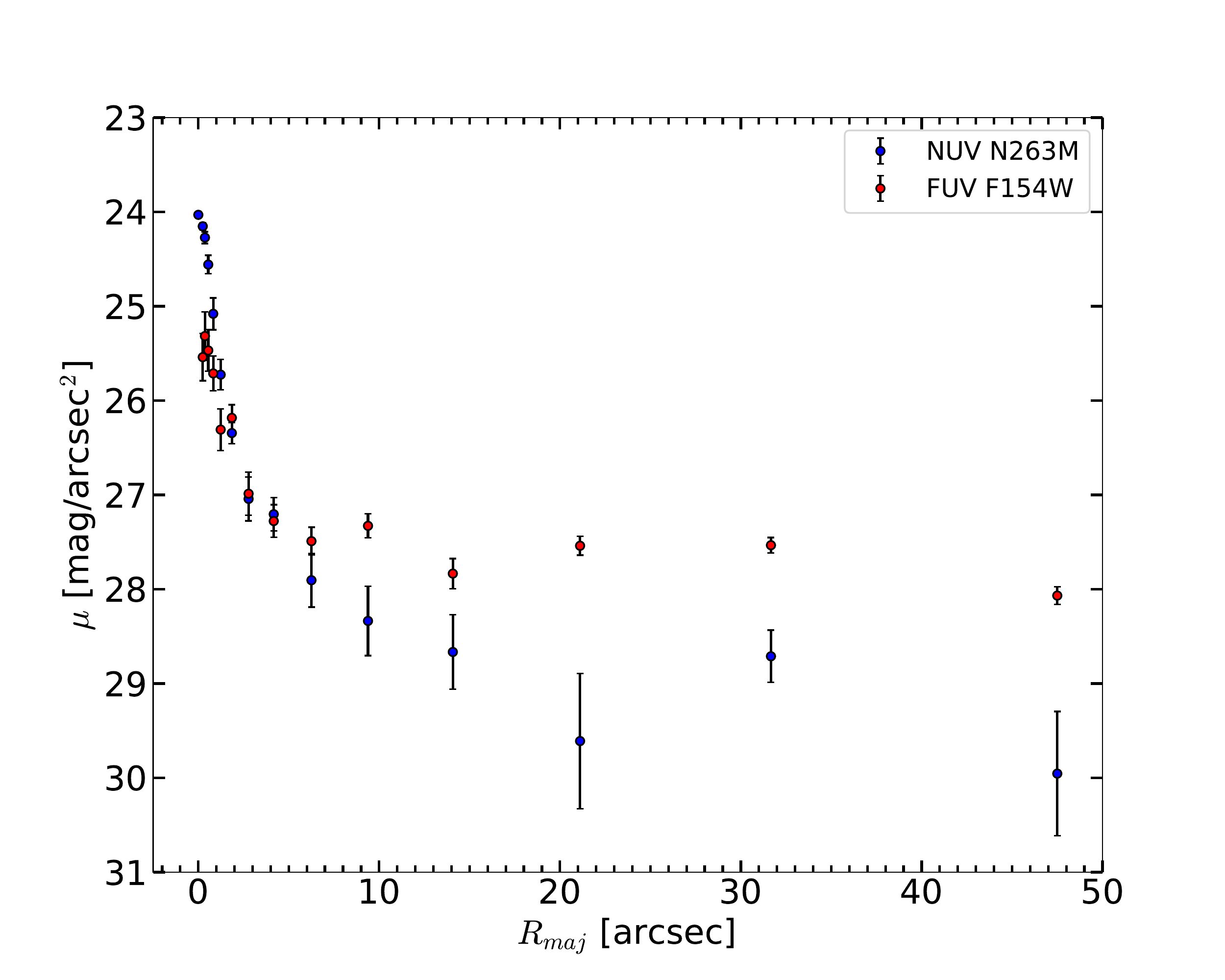}
  \caption{Surface brightness profiles in near and far-UV bands of AstroSat/UVIT. The galaxy has extremely low surface brightness level outside the central region. }
  \label{fig:UVSBP}
\end{flushleft}
\end{figure}

In Fig.~\ref{fig:UVSBP}, we show the surface brightness profile (SBP) of Malin 1 in F154W and N263M bands. Since the galaxy is extremely diffuse in FUV and NUV, we have not run the IRAF ellipse task to derive its SBPs. Instead, the SBPs are derived by placing concentric circular apertures on the galaxy and reading the counts within each aperture. Both the F154W and N263M SBPs are corrected for the foreground dust extinction as mentioned in Sec.~\ref{sec:dust}. It is useful to compare our UV SBPs with that from GALEX. At 10" from the centre, the F154W surface brightness level is $\sim 27.5$~mag~arcsec$^{-2}$ and in N263M band, it is about $\sim 28.4$~mag~arcsec$^{-2}$. At a radius of 20", the N263M SB level is $\sim 29.5$~mag~arcsec$^{-2}$ while the FUV SBP stays flat at $\sim 28$~mag~arcsec$^{-2}$ except at the location of the spiral arms. When comparing these with the previous studies \citep{Boissier2016}, we find that the UVIT probes deeper in surface brightness level at 10" radius while at 20" the N263M filter goes $1$~mag~arcsec$^{-2}$ deeper than GALEX NUV. At smaller radii i.e., within 10" radius, the surface brightness level rises steeply both in FUV and NUV and this has become possible because of the better image resolution in UVIT. From the UV surface brightness profiles, we find that the $FUV-NUV \simeq 0$ in the central region consistent with previous result \citep{Boissier2016}. However, at radii $> 10"$, the FUV SBP flattens out and $FUV-NUV \simeq -1$ - indicating young, bluer stellar population in the outskirts. Since Malin 1 is interacting with two other galaxies - one is in-spiraling i.e., Malin~1B and another, SDSS J123708.91+142253.2 \citep{Reshetnikovetal2010}, currently at a larger distance, might be responsible for the excitation of star-formation activity in the overall galaxy and might have elevated the FUV flux in the whole galaxy. This is not so unlikely as the galaxy Malin~1 has a huge reservoir of neutral hydrogen gas. 
\par
\begin{figure}
\includegraphics[height=.33\textheight]{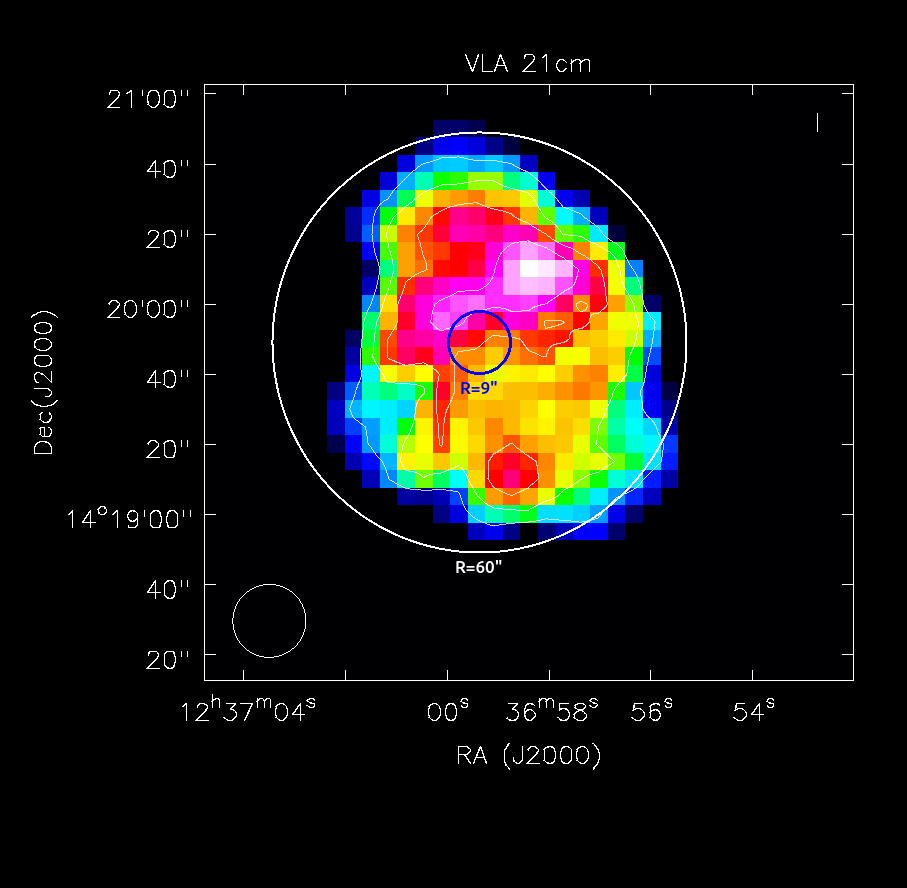}
\caption{HI 21 cm map of Malin 1 (reproduced here with permission from T. Pickering). The HI extends up to 60" (white circle). The blue circle corresponds to the size of Malin 1's central 9" region. A rough estimate of the HI mass within the blue circle could be $M_{HI} = 2.6 \times 10^9$~M$_{\odot}$. \cite{Pickeringetal1997} find evidence for a strong warp, which may explain all or part of the apparent non-axisymmetric HI distribution seen here.}
\label{fig:HI}
\end{figure}

The most validated 21~cm emission from VLA observations of Malin 1 by \cite{Pickeringetal1997} shows an extended H{{\sc i}} disk up to $\sim 60"$ radius (see Fig.~\ref{fig:HI}). The total H{{\sc i}} mass of this huge disk was estimated by them to be $6.8 \times 10^{10}$~M$_{\odot}$ corresponding to an integrated flux of $2.5$~Jy~km/s (with a H$_0$ = 75 km s$^{-1}$ Mpc$^{-1}$). 
They also found a slowly rising rotation curve that flattened to 210 km s$^{-1}$ at $\sim$ 40". However, \cite{Lellietal2010} note that the slow rise in the central region could well be an effect of beam smearing (which affects only the central parts). On a re-analysis of \cite{Pickeringetal1997} data with a technique that reduces beam smearing, they find that the circular velocity touches 237 km s$^{-1}$ at 8" itself. This implies that Malin 1 has a steeply rising rotation curve, which is a unique feature of an HSB disk. Based on this, \cite{Lellietal2010} argue that Malin 1 has an inner early-type HSB disk followed by an extended LSB disk. As mentioned earlier, the outcome from the rotation curve analysis is in accordance with findings from HST observation which describes the central region of Malin 1 as a normal SB0/a galaxy surrounded by a huge LSB disk \citep{Barth2007}. 
\par
As evident in Fig.~\ref{fig:HI}, the apparent distribution in the extended H{{\sc i}} disk is far from centrally concentrated. \cite{Pickeringetal1997} conclude that the H{{\sc i}} disk is very strongly warped from the twisted contours seen in their individual channel maps. A strong northern warp (towards us) coupled with the inclination of 38$^\circ$ may well explain the observed non-axisymmetry in H{{\sc i}} distribution. Yet, it is unclear whether the warp alone is sufficient to explain it or the disk is also lopsided to an extent which could be a result of the ongoing tidal-interaction \citep{YozinBekki2014,vanEymerenetal2011}. From Fig.~\ref{fig:HI}, it also clear that the central 9" region which harbours a bar has a lopsided H{\sc i} distribution. This is interesting as well as exciting given the morphology of the central region. Normally, galaxies with S0 morphology are gas-poor \citep{Chamarauxetal1986,Haynesetal1990,Mastersetal2012} and if there is a bar, one expects to find H{{\sc i}} holes \citep{vanDriel1991,Athanassoulaetal2013,Newnhametal2020}. In the following section, we discuss the implication and possible connection between the bar, far-UV emission and the H{{\sc i}} observation. 

\begin{figure}
\begin{flushleft}
\includegraphics[height=.78\textheight]{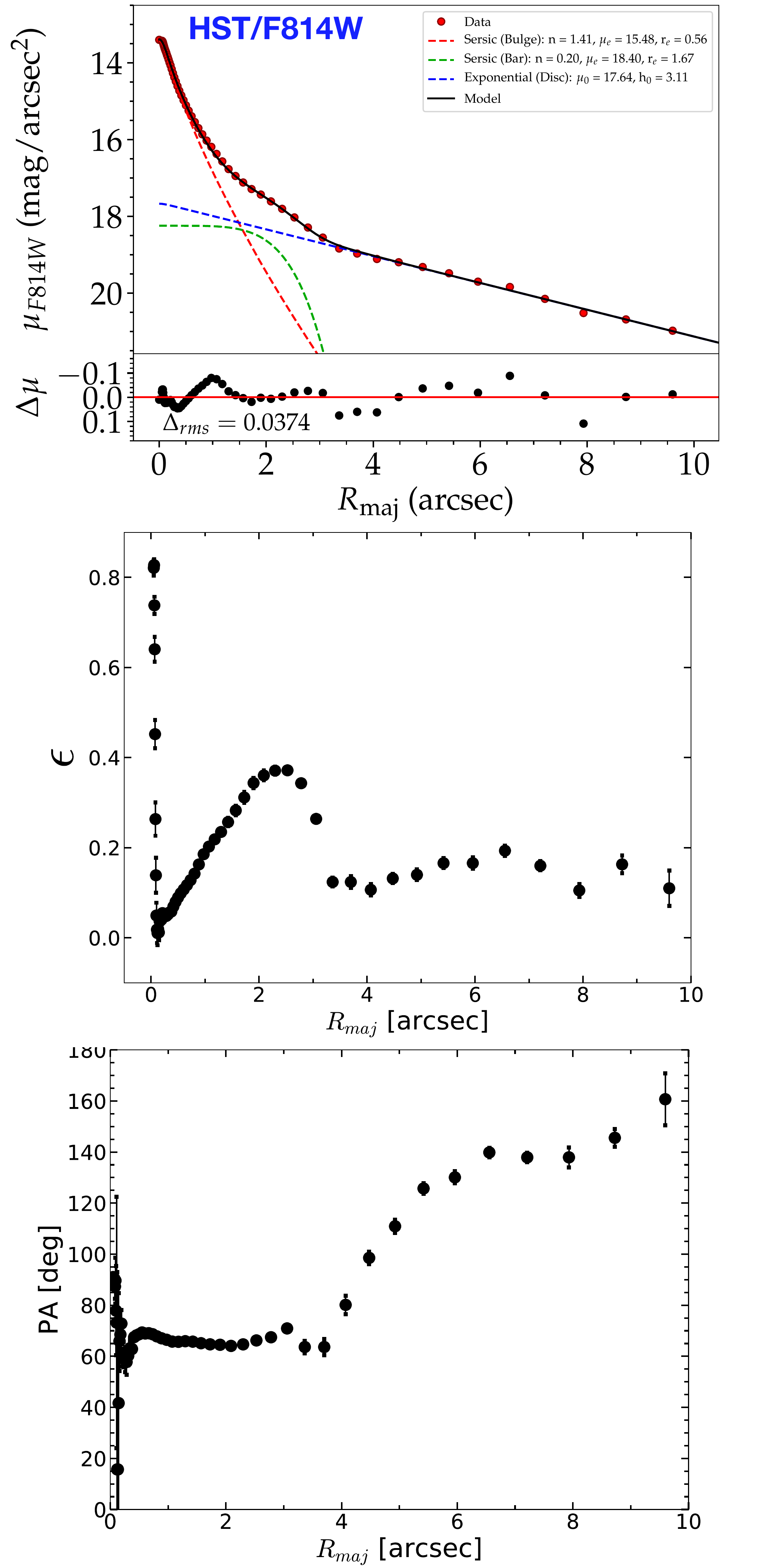}
\caption{The top panel shows the HST F814W surface brightness profile fitted with two Sersic components: one for the bulge (dashed orange line), and the other for the bar (dashed green line). The inner exponential profile is indicated by the dashed blue line. Black solid line is the best fit model. The middle and lower panels show the change in ellipticity and the position angle respectively.}
\label{fig:centralSBP}
\end{flushleft}
\end{figure}

\begin{figure*}
\begin{flushleft}
  \includegraphics[height=.27\textheight]{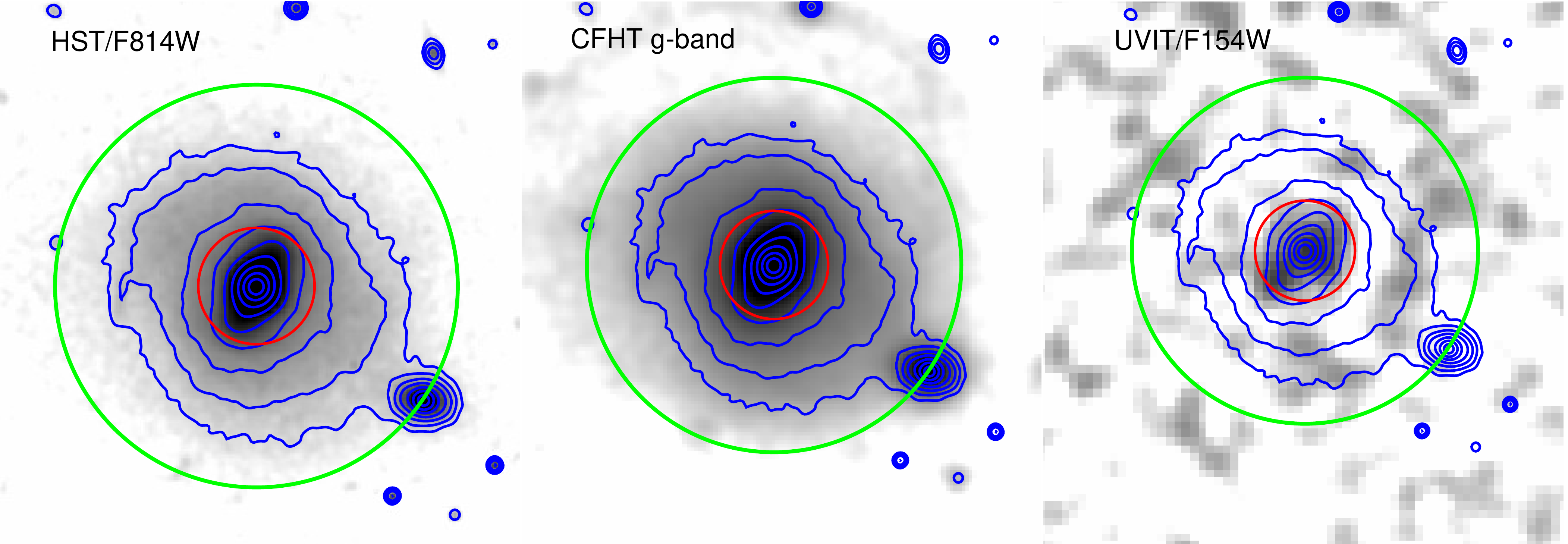}
  \caption{Central morphology of Malin 1 - multi-wavelength view from HST I-band to AstroSat/UVIT far-UV. A bar of size 2.6" is visible in the HST/F814W band. The radius of the green circle is 9" (which contains the inner exponential disk). F814W contours are overlayed on the rest of the images. There is FUV emission from the bar region. The brightest pixel in F154W within the bar region is 0.00051 ct/s.}
  \label{fig:central}
\end{flushleft}
\end{figure*}

\section{Central disk and the bar}
\label{sec:bar}

The high resolution HST image has revealed a central bar \citep{Barth2007} in the heart of Malin~1 within 9" radius. Here for the sake of completeness, we model the surface brightness profile of the central 9" region observed in HST/F814W band (see Fig.~\ref{fig:centralSBP}). The central region has three components - a bulge, a bar and an exponential disk. We have used profiler \citep{Ciambur16} to model the inner surface brightness profile taking into account the HST PSF. The bulge is a compact pseudobulge with sersic index $n=1.4$ and $r_{e}=0.56"=0.87$~kpc in the F814W band. The exponential disk has a central surface brightness of $\mu_{0}=17.64$~mag~arcsec$^{-2}$ and a scale-length of 4.8~kpc (in F814W). This disk is clearly a high surface brightness disk like our own Galaxy and our analysis confirms previously findings in this regard \cite{Barth2007,Lellietal2010, Boissier2016}. 
The lower panels of Fig.~\ref{fig:centralSBP} show the radial variation of the ellipticity and position angle (PA) of the central region. The ellipticity has a peak at around 2.6"(=4~kpc) which is considered here as the radius of the bar \citep{Erwin2005}. Beyond this region i.e., outside the bar ($ \sim 3" - 5"$), the PA changes abruptly and there is a hint of faint spiral arm (on the East-ward direction, see the CFHT u-band image in Fig.~\ref{fig:cfht-UVIT}) which might have been driven by the bar itself \citep{Sahaetal2010,SahaElmegreen2016}. Otherwise, the inner part resembles S0 galaxies with a bar and pseudo-bulge \citep{Vaghmareetal2015,Vaghmareetal2018}. If we consider the optical color $g-r =1.16$ for this 9" central region, we find the mass-to-light ratio is 9.1 following \cite{BelldeJong2001} and a stellar mass of $M_{*} =8.9\times 10^{11} M_{\odot}$ - this is about $17$ times more massive than the Milky Way's stellar mass, although the size is similar. Using our rough estimate of the H{{\sc i}} mass within this region, we find $M_{HI}/M_{*}=0.003$ - pushing the central region of Malin 1 close to one of the most gas-poor galaxies \citep{Haynesetal1990,Mastersetal2012}. Note that the galaxy contains a significant amount of cold H{{\sc i}} gas but because of the huge stellar mass ($\sim 1.8\times 10^{12} M_{\odot}$), the overall gas fraction is only about $\sim 4\%$. 

Fig.~\ref{fig:central} shows the central morphology of Malin~1 in HST/F814W, CFHT g-band, and UVIT/F154W bands. The bar is most prominent in the HST F814W but not so prominent in the CFHT optical images. However, as we saw earlier, the CFHT $g-i$ colormap reveals the bar mostly clearly (see Fig.~\ref{fig:colormap}). In Fig.~\ref{fig:central}, we also show the far-UV emission morphology in the central region alongside other bands. The central region has a few clumps in far-UV suggesting young, recent star-formation. But most of the FUV emission is originating from the bar region. The optical contours and the position angle show that the bar is oriented along the south-east to north-west direction with a position angle of $63^{\circ}$ for the optical bar. We also see emission from the bar region in the near-UV N263M band (see the bottom right panel of Fig.~\ref{fig:GALEX-UVIT}). The background subtracted far-UV flux within the bar region is $S=0.008775$~ct/s corresponding to $22.92$~AB mag. After the aperture and foreground dust extinction correction using \citet{Schlegel98}, the bar region has $22.64$~AB mag in F154W band. In the N263M band, the background subtracted flux within the bar region (same aperture) is $0.018219$~ct/s corresponding to $22.53$~AB mag which after aperture and foreground correction becomes $22.28$~AB mag. Within the same aperture (of 2.6" radius), we have computed the bar magnitudes in CFHT u,g, and i bands and they are 19.65, 17.98 and 16.69 AB mag respectively (all mags are foreground dust corrected). The FUV-NUV (F154W-N263M) color of the bar region is 0.36 - suggesting extremely blue and the presence of hot, young O, OB stars. While the $g-i=1.29$ color of the bar suggest red and older stellar population. The same is reflected in other UV-optical colours such as FUV-i=5.7, NUV-g=4.27 - indicating recent star formation activity \citep{Schawinski09} in the bar region. The broad-band colours and the central strong emission lines such as H$\alpha$ and [O{{\sc ii}}] suggest mixed stellar population with very young stars. Contrary to quenching due to the bar \citep{Georgeetal2019}, we see rejuvenation of the bar \citep{BarwaySaha2020} probably induced by the companion galaxies but this remains to be investigated with more detailed IFU-like observation. Overall, it appears that the gigantic Malin~1 has a massive Milky Way size stellar disk at its heart.

\subsection{Central Star Formation activity}
\label{sec:SFR}
The central region of Malin~1 shows both absorption and emission lines. The strong absorption lines are due to metals in the stellar atmospheres of mostly low-luminosity stellar population. It has several absorption lines e.g., Mg, Na, G-band, Ca(H/K) - suggestive of metal rich environment or old evolved stars.  We use the O3N2 relation \citep{PettiniPagel2004} to estimate the gas-phase Oxygen abundance. we find the metallicity of the galaxy to be 8.69 which is close to solar metallicity. The emission lines such H$\alpha$, [O{{\sc ii}}], [O{{\sc iii}}] suggest recent star formation activity. In fact, there are scattered recent star-formation activity in the entire disk all the way up to about 60" as revealed by our UVIT image in F154W band, especially along the spiral arms. This is evident upon close inspection of the CFHT g-band and the F154W image - where one can trace spiral arms and FUV clumps; also clear from the $g-i$ color map (Fig.~\ref{fig:colormap}). 
\par
We estimate the central star formation rate using the SDSS spectroscopic data which provides line fluxes for the central 3" region. We use the \cite{Kennicutt98} relation
 
\begin{equation}
SFR~(M_{\odot}yr^{-1}) = 7.93\times 10^{-42} L_{H\alpha} ~ (erg~s^{-1}),
\end{equation}

to estimate the SFR within central 3" region (which covers the middle part of the bar) from the Balmer corrected H$\alpha$ flux $336.7 \pm 8.7 \times 10^{-17}$ ~erg~s$^{-1}$~cm$^{-2}$ (see Sec.~\ref{sec:dust}). Then the dust corrected H$\alpha$ luminosity is $5.74 \pm 0.14 \times 10^{40}$~erg~s$^{-1}$. The estimated central SFR is found to be 0.45$\pm$0.01 M$_{\odot}$yr$^{-1}$. 

\subsection{Far-UV SFR}

Central SFR within 2.6 arcsec is calculated using the FUV  magnitude. We correct the observed FUV magnitude of the bar region (see Sec.~\ref{sec:bar}) for the internal dust extinction using Balmer decrement method and following \cite{Calzettietal2000} dust attenuation law at $\lambda_{mean}=1541 \AA$. The dust corrected (both internal and foreground) FUV magnitude of the bar region is $22.4$ AB mag. The corresponding flux from the bar region is $5.0 \pm 1.6\times 10^{-17}$~erg~s$^{-1}$~cm$^{-2}$~$\AA^{-1}$ ($3.9 \pm 1.2 \times 10^{-29}$~erg~s$^{-1}$~cm$^{-2}$~Hz$^{-1}$) and luminosity = $6.74 \pm 2.1 \times 10^{26}$~erg~s$^{-1}$~Hz$^{-1}$. Further we estimate the far-UV SFR using the following relation \citep{Kennicutt98}:

\begin{equation}
SFR~(M_{\odot}yr^{-1}) = 1.4\times 10^{-28} L_{FUV} ~ (erg~s^{-1}~Hz^{-1})
\end{equation}

\noindent The estimated FUV SFR within the central 2.6" (i.e., the bar region) is $0.094 \pm 0.03$~M$_{\odot}yr^{-1}$. In deriving the above relation, \citet{Kennicutt98} uses the Salpeter Initial Mass Function (IMF) with mass limits of 0.1 to 100 M$_{\odot}$ and stellar  population models with solar abundance. 

We also estimate the SFR within 2.6 arcsec using the recent empirical relation by \cite{KarachentsevKaisina2013}:

\begin{equation}
\log SFR~(M_{\odot}yr^{-1}) = 2.78 - 0.4 \times m_{FUV} + 2\log(D),
\end{equation}

\noindent where D is in Mpc and $m_{FUV}$ refers to the FUV magnitude. The estimated FUV SFR is $0.094 \pm 0.008$~M$_{\odot}yr^{-1}$ in good agreement with that obtained using \citep{Kennicutt98}. The specific starformation rate $sSFR = 1.00 \times 10^{-13}$~yr$^{-1}$ - placing the central region of the galaxy in the quenched category marked by $\log{sSFR}=-11.8$ \citep{BarwaySaha2020}. Even if the star formation rate jumps by a factor of 10, the central region of Malin~1 would be classified as quenched. Nevertheless, the detailed process of quenching remains unclear \citep{vandenBoschetal2008,Martigetal2009,PengMaiolinoCochrane2015}. The strong emission lines from the central region \citep{Junaisetal2020} and our UV observation suggest that there is recent star-formation activity and the presence of hot blue stars. Together, we seem to arrive at a complex scenario of the central region of Malin~1. Whatever might the case, it would be interesting to investigate in detail the central region of this galaxy.

\section{Discussion and Conclusions}
\label{sec:conclusion}
Malin 1 is a classic example of a giant low-surface brightness galaxy \citep{ImpeyBothun1989, MooreParker06, Galazetal2015,Boissier2016} which is currently undergoing tidal interaction with two external galaxies \citep{Reshetnikovetal2010}. The central region (i.e., within 14 kpc) of Malin 1 has a HSB stellar disk with a compact, central pseudobulge, and a bar. The bar has a radius of 4.~kpc. This central region is equivalent to $\sim 17$ Milky Way size stellar disks put together within this 9" region. The central disk has a few FUV clumps ($> S/N=3$) - indicating hot, young stars but only a hint of faint spiral structure. So the emerging picture of Malin 1 is that at the heart it, has a star-forming HSB Milky Way size stellar disk surrounded by an intermediate LSB exponential disk extending upto about 30 kpc. Beyond this lies the extremely low-surface brightness stellar envelope. The apparently simpler giant LSB galaxy has a complex layered structure - might hold important clues to the true complexity of galaxy formation yet to be unfolded \citep{Coleetal2000,Martinetal2019,Kulieretal2020}.

The star-forming bar in Malin~1 is surprising because of the central S0-like morphology. Our UVIT observation shows FUV emission from the bar region - indicating ongoing star formation. In addition, the SDSS central 3" fibre shows strong emission lines such as H$\alpha$ and [O{{\sc ii}}] indicating again the most recent (maybe ~10 Myr old) star formation activity in the bar; a similar scenario has been observed in NGC 2903 \citep{Georgeetal2020}. The bar appears to play a dual role in the star-formation activity in galaxies. It can funnel gas inwards \citep{Combes2004} and ignite star-formation activity in the central region of a galaxy and lead to the formation of a pseudobulge \citep{Athanassoula1992,KK2004,Jogeeetal2005,Linetal2017}. On the other hand, a bar can also lead to the star-formation quenching \citep{Jamesetal2009,JamesPercival2016,Khoperskovetal2018,Georgeetal2019,Georgeetal2020}. The bar in Malin~1 has induced star-formation, probably due to the ongoing interaction; especially due to the in-spiraling Malin 1B. In addition, the strong lopsidedness of the atomic hydrogen gas might also lead to inward movement of gas due to outward angular momentum transport \citep{Sahajog2014}.

Our primary conclusions from this work are:

\begin{itemize}
    \item 
    Using the high resolution UVIT observation, we show that there is scattered star formation all over the disk up to about 93 kpc. The FUV emission is clearly seen to follow the large scale spiral arms of Malin 1.
    \item
    We perform multi-component decomposition of the light profile of Malin 1 in - HST/F814W and CFHT g-band. The CFHT deep image in the g-band reveals that Malin 1 has a very faint low-surface brightness outer disk, and an intermediate LSB disk. Both disks follow exponential profile - the outer one has a scale-length of 47~kpc while the intermediate one has a scale-length of 3.9~kpc. 
    \item
    The central 14~kpc region is modelled using F814W bands. Our decomposition shows that the central region has three components namely a compact pseudobulge, a bar and a HSB exponential stellar disk. 
    
    \item
      The high resolution UVIT imaging data shows FUV emission from the bar region - indicating ongoing star-formation activity. The FUV-NUV colour of the bar is 0.36 - indicating the presence of hot, young stars. Based on the extinction corrected FUV flux of the bar region, we estimate an SFR$=0.09$~M$_{\odot}$~yr$^{-1}$. Whereas the SFR based on the central H$\alpha$ line flux is $=0.45$~M$_{\odot}$~yr$^{-1}$. 
      
      \item The better PSF and sensitivity of UVIT have allowed us to detect FUV fluxes within the Malin 1B. The galaxy Malin 1B has a magnitude of $24.9\pm 0.32$ AB mag.
      
     \item
     The surface brightness profiles in FUV and NUV upto about 77~kpc shows that FUV is flatter than NUV in the outer region - suggesting bluer population and scattered star-formation till the end of the disk.

\end{itemize}

\section*{Acknowledgements}
The AstroSat/UVIT observation of Malin 1 and the preliminary data analysis was done as part of the collaborative program under bilateral grant DST/INT/South Africa/P-03/2016 of the Indo-South Africa Flagship Program in Astronomy. A good part of this work was initiated during the visit of KS to South African Astronomical Observatory, Cape Town. KS and SB acknowledge the generous financial support under the same. The UVIT project is a collaboration between IIA, IUCAA, TIFR, ISRO from India and CSA from Canada. This publication uses the UVIT data obtained from Indian Space Science Data Centre (ISSDC)of ISRO, where the data for AstroSat mission is archived. We thank the anonymous referee for an insightful comments which improved the quality of the paper. We thank T. Pickering for kindly making the HI data available to us. SD gratefully acknowledges the support from Department of Science and Technology (DST), New Delhi under the INSPIRE faculty  Scheme (sanctioned No: DST/INSPIRE/04/2015/000108).



\begin{thebibliography}{}
\expandafter\ifx\csname natexlab\endcsname\relax\def\natexlab#1{#1}\fi

\bibitem[{{Aguerri} {$et~al$.}(2001){Aguerri}, {Balcells}, \&
  {Peletier}}]{Aguerrietal2001}
{Aguerri}, J.~A.~L., {Balcells}, M., \& {Peletier}, R.~F. 2001, \aap, 367, 428

\bibitem[{{Athanassoula}(1992)}]{Athanassoula1992}
{Athanassoula}, E. 1992, \mnras, 259, 345

\bibitem[{{Athanassoula} {$et~al$.}(2013){Athanassoula}, {Machado}, \&
  {Rodionov}}]{Athanassoulaetal2013}
{Athanassoula}, E., {Machado}, R. E.~G., \& {Rodionov}, S.~A. 2013, \mnras,
  429, 1949

\bibitem[{{Barth}(2007)}]{Barth2007}
{Barth}, A.~J. 2007, \aj, 133, 1085

\bibitem[{{Barway} \& {Saha}(2020)}]{BarwaySaha2020}
{Barway}, S., \& {Saha}, K. 2020, \mnras, 495, 4548

\bibitem[{{Bell} \& {de Jong}(2001)}]{BelldeJong2001}
{Bell}, E.~F., \& {de Jong}, R.~S. 2001, \apj, 550, 212

\bibitem[{{Bertin} \& {Arnouts}(1996)}]{BertinArnouts1996}
{Bertin}, E., \& {Arnouts}, S. 1996, \aaps, 117, 393

\bibitem[{{Boissier} {$et~al$.}(2016){Boissier}, {Boselli}, {Ferrarese},
  {C{\^o}t{\'e}}, {Roehlly}, {Gwyn}, {Cuillandre}, {Roediger}, {Koda},
  {Mu{\~n}os Mateos}, {Gil de Paz}, \& {Madore}}]{Boissier2016}
{Boissier}, S., {Boselli}, A., {Ferrarese}, L., {$et~al$.} 2016, \aap, 593,
  A126

\bibitem[{{Bothun} {$et~al$.}(1987){Bothun}, {Impey}, {Malin}, \&
  {Mould}}]{Bothunetal1987}
{Bothun}, G.~D., {Impey}, C.~D., {Malin}, D.~F., \& {Mould}, J.~R. 1987, \aj,
  94, 23

\bibitem[{{Braine} {$et~al$.}(2000){Braine}, {Herpin}, \&
  {Radford}}]{Braineetal2000}
{Braine}, J., {Herpin}, F., \& {Radford}, S.~J.~E. 2000, \aap, 358, 494

\bibitem[{{Calzetti} {$et~al$.}(2000) {Armus}, {Bohlin}, {$et~al.$}}]{Calzettietal2000} {Calzetti}, D., {Armus}, L., {Bohlin}, R.~C., {$et~al.$}  2000, \apj, 533, 682


\bibitem[{{Chamaraux} {$et~al$.}(1986){Chamaraux}, {Balkowski}, \&
  {Fontanelli}}]{Chamarauxetal1986}
{Chamaraux}, P., {Balkowski}, C., \& {Fontanelli}, P. 1986, \aap, 165, 15

\bibitem[{{Ciambur}(2016)}]{Ciambur16}
{Ciambur}, B.~C. 2016, \pasa, 33, e062

\bibitem[{{Cole} {$et~al$.}(2000){Cole}, {Lacey}, {Baugh}, \&
  {Frenk}}]{Coleetal2000}
{Cole}, S., {Lacey}, C.~G., {Baugh}, C.~M., \& {Frenk}, C.~S. 2000, \mnras,
  319, 168

\bibitem[{{Combes}(2004)}]{Combes2004}
{Combes}, F. 2004, IAUS, 222, 383


\bibitem[{{Das} {$et~al$.}(2006){Das}, {O'Neil}, {Vogel}, \&
  {McGaugh}}]{Dasetal2006}
{Das}, M., {O'Neil}, K., {Vogel}, S.~N., \& {McGaugh}, S. 2006, \apj, 651, 853

\bibitem[{{Eliche-Moral} {$et~al$.}(2018){Eliche-Moral},
  {Rodr{\'\i}guez-P{\'e}rez}, {Borlaff}, {Querejeta}, \&
  {Tapia}}]{Eliche-Moraletal2018}
{Eliche-Moral}, M.~C., {Rodr{\'\i}guez-P{\'e}rez}, C., {Borlaff}, A.,
  {Querejeta}, M., \& {Tapia}, T. 2018, \aap, 617, A113

\bibitem[{{Erwin}(2005)}]{Erwin2005}
{Erwin}, P. 2005, \mnras, 364, 283

\bibitem[{{Ferrarese} {$et~al$.}(2012){Ferrarese}, {C{\^o}t{\'e}}, {Cuilland
  re}, {Gwyn}, {Peng}, {MacArthur}, {Duc}, {Boselli}, {Mei}, {Erben},
  {McConnachie}, {Durrell}, {Mihos}, {Jord{\'a}n}, {Lan{\c{c}}on}, {Puzia},
  {Emsellem}, {Balogh}, {Blakeslee}, {van Waerbeke}, {Gavazzi}, {Vollmer},
  {Kavelaars}, {Woods}, {Ball}, {Boissier}, {Courteau}, {Ferriere}, {Gavazzi},
  {Hildebrandt}, {Hudelot}, {Huertas-Company}, {Liu}, {McLaughlin}, {Mellier},
  {Milkeraitis}, {Schade}, {Balkowski}, {Bournaud}, {Carlberg}, {Chapman},
  {Hoekstra}, {Peng}, {Sawicki}, {Simard}, {Taylor}, {Tully}, {van Driel},
  {Wilson}, {Burdullis}, {Mahoney}, \& {Manset}}]{Ferrareseetal2012}
{Ferrarese}, L., {C{\^o}t{\'e}}, P., {Cuilland re}, J.-C., {$et~al$.} 2012,
  \apjs, 200, 4

\bibitem[{{Freeman}(1970)}]{Freeman1970}
{Freeman}, K.~C. 1970, \apj, 160, 811

\bibitem[{{George} {$et~al$.}(2019){George}, {Joseph}, {Mondal}, {Subramanian},
  {Subramaniam}, \& {Paul}}]{Georgeetal2019}
{George}, K., {Joseph}, P., {Mondal}, C., {$et~al$.} 2019, \aap, 621, L4

\bibitem[{{George} {$et~al$.}(2020){George}, {Joseph}, {Mondal}, {Subramanian},
  {Subramaniam}, \& {Paul}}]{Georgeetal2020}
{George}, K., {Joseph}, P., {Mondal}, C., {$et~al$.} 2020, \aap, 644A, 79

\bibitem[{{Galaz} {$et~al$.}(2015){Galaz}, {Milovic}, {Suc}, {Busta},
	{Lizana},{Infante} \& {Royo}}]{Galazetal2015}
{Galaz}, G., {Milovic}, C., {Suc}, V., {$et~al$.} 2015, \apjl, 815, L29


\bibitem[{{Gouliermis} {$et~al$.}(2017){Gouliermis}, {Elmegreen}, {Elmegreen},
  {Calzetti}, {Cignoni}, {Gallagher}, {Kennicutt}, {Klessen}, {Sabbi},
  {Thilker}, {Ubeda}, {Aloisi}, {Adamo}, {Cook}, {Dale}, {Grasha}, {Grebel},
  {Johnson}, {Sacchi}, {Shabani}, {Smith}, \& {Wofford}}]{Gouliermisetal2017}
{Gouliermis}, D.~A., {Elmegreen}, B.~G., {Elmegreen}, D.~M., {$et~al$.} 2017,
  \mnras, 468, 509

\bibitem[{{Haynes} {$et~al$.}(1990){Haynes}, {Herter}, {Barton}, \&
  {Benensohn}}]{Haynesetal1990}
{Haynes}, M.~P., {Herter}, T., {Barton}, A.~S., \& {Benensohn}, J.~S. 1990,
  \aj, 99, 1740

\bibitem[{{Impey} \& {Bothun}(1989)}]{ImpeyBothun1989}
{Impey}, C., \& {Bothun}, G. 1989, \apj, 341, 89

\bibitem[{{James} {$et~al$.}(2009){James}, {Bretherton}, \&
  {Knapen}}]{Jamesetal2009}
{James}, P.~A., {Bretherton}, C.~F., \& {Knapen}, J.~H. 2009, \aap, 501, 207

\bibitem[{{James} \& {Percival}(2016)}]{JamesPercival2016}
{James}, P.~A., \& {Percival}, S.~M. 2016, \mnras, 457, 917

\bibitem[{{Jogee} {$et~al$.}(2005){Jogee}, {Scoville}, \&
  {Kenney}}]{Jogeeetal2005}
{Jogee}, S., {Scoville}, N., \& {Kenney}, J. D.~P. 2005, \apj, 630, 837

\bibitem[{{Junais} {$et~al$.}(2020){Junais}, {Boissier}, {Epinat}, {Amram},
  {Madore}, {Boselli}, {Koda}, {Gil de Paz}, {Mu{\~n}os Mateos}, \&
  {Chemin}}]{Junaisetal2020}
{Junais}, {Boissier}, S., {Epinat}, B., {$et~al$.} 2020, \aap, 637, A21

\bibitem[{{Karachentsev} \& {Kaisina}(2013)}]{KarachentsevKaisina2013}
{Karachentsev}, I.~D., \& {Kaisina}, E.~I. 2013, \aj, 146, 46

\bibitem[{{Kennicutt}(1998)}]{Kennicutt98}
{Kennicutt}, Robert~C., J. 1998, \apj, 498, 541

\bibitem[{{Khoperskov} {$et~al$.}(2018){Khoperskov}, {Haywood}, {Di Matteo},
  {Lehnert}, \& {Combes}}]{Khoperskovetal2018}
{Khoperskov}, S., {Haywood}, M., {Di Matteo}, P., {Lehnert}, M.~D., \&
  {Combes}, F. 2018, \aap, 609, A60


\bibitem[{{Kulier} {$et~al$.}(2020){Kulier}, {Galaz}, {Padilla}, \&
  {Trayford}}]{Kulieretal2020}
{Kulier}, A., {Galaz}, G., {Padilla}, N.~D., \& {Trayford}, J.~W. 2020, \mnras,
  496, 3996

\bibitem[{{Lang} {$et~al$.}(2014){Lang}, {Holley-Bockelmann}, \&
  {Sinha}}]{Langetal2014}
{Lang}, M., {Holley-Bockelmann}, K., \& {Sinha}, M. 2014, \apjl, 790, L33

\bibitem[{{Lelli} {$et~al$.}(2010){Lelli}, {Fraternali}, \&
  {Sancisi}}]{Lellietal2010}
{Lelli}, F., {Fraternali}, F., \& {Sancisi}, R. 2010, \aap, 516, A11

\bibitem[{{Lin} {$et~al$.}(2017){Lin}, {Li}, {He}, {Xiao}, \&
  {Wang}}]{Linetal2017}
{Lin}, L., {Li}, C., {He}, Y., {Xiao}, T., \& {Wang}, E. 2017, \apj, 838, 105

\bibitem[{{{\L}okas} {$et~al$.}(2014){{\L}okas}, {Athanassoula}, {Debattista},
  {Valluri}, {Pino}, {Semczuk}, {Gajda}, \& {Kowalczyk}}]{Lokasetal2014}
{{\L}okas}, E.~L., {Athanassoula}, E., {Debattista}, V.~P., {$et~al$.} 2014,
  \mnras, 445, 1339

\bibitem[{{Martig} {$et~al$.}(2009){Martig}, {Bournaud}, {Teyssier}, \&
  {Dekel}}]{Martigetal2009}
{Martig}, M., {Bournaud}, F., {Teyssier}, R., \& {Dekel}, A. 2009, \apj, 707,
  250

\bibitem[{{Martin} {$et~al$.}(2019){Martin}, {Kaviraj}, {Laigle}, {Devriendt},
  {Jackson}, {Peirani}, {Dubois}, {Pichon}, \& {Slyz}}]{Martinetal2019}
{Martin}, G., {Kaviraj}, S., {Laigle}, C., {$et~al$.} 2019, \mnras, 485, 796

\bibitem[{{Masters} {$et~al$.}(2012){Masters}, {Nichol}, {Haynes}, {Keel},
  {Lintott}, {Simmons}, {Skibba}, {Bamford}, {Giovanelli}, \&
  {Schawinski}}]{Mastersetal2012}
{Masters}, K.~L., {Nichol}, R.~C., {Haynes}, M.~P., {$et~al$.} 2012, \mnras,
  424, 2180

\bibitem[{{Miwa} \& {Noguchi}(1998)}]{MiwaNoguchi1998}
{Miwa}, T., \& {Noguchi}, M. 1998, \apj, 499, 149

\bibitem[{{Moore} \& {Parker}(2006)}]{MooreParker06}
{Moore}, L., \& {Parker}, Q.~A. 2006, \pasa, 23, 165

\bibitem[{{Morrissey} {$et~al$.}(2007){Morrissey}, {Conrow}, {Barlow}, {Small},
  {Seibert}, {Wyder}, {Budav{\'a}ri}, {Arnouts}, {Friedman}, {Forster},
  {Martin}, {Neff}, {Schiminovich}, {Bianchi}, {Donas}, {Heckman}, {Lee},
  {Madore}, {Milliard}, {Rich}, {Szalay}, {Welsh}, \& {Yi}}]{Morrisseyetal2007}
{Morrissey}, P., {Conrow}, T., {Barlow}, T.~A., {$et~al$.} 2007, \apjs, 173,
  682


\bibitem[{{Osterbrock} \& {Ferland}(2006)}]{OsterbrockFerland2006}
{Osterbrock}, D.~E., \& {Ferland}, G.~J. 2006, {Astrophysics of gaseous nebulae
  and active galactic nuclei}

\bibitem[{{Pahwa} \& {Saha}(2018)}]{PahwaSaha2018}
{Pahwa}, I., \& {Saha}, K. 2018, \mnras, 478, 4657

\bibitem[{{Peng} {$et~al$.}(2015){Peng}, {Maiolino}, \&
  {Cochrane}}]{PengMaiolinoCochrane2015}
{Peng}, Y., {Maiolino}, R., \& {Cochrane}, R. 2015, \nat, 521, 192

\bibitem[{{Pettini} \& {Pagel}(2004)}]{PettiniPagel2004}
{Pettini}, M., \& {Pagel}, B. E.~J. 2004, \mnras, 348, L59

\bibitem[{{Pickering} {$et~al$.}(1997){Pickering}, {Impey}, {van Gorkom}, \&
  {Bothun}}]{Pickeringetal1997}
{Pickering}, T.~E., {Impey}, C.~D., {van Gorkom}, J.~H., \& {Bothun}, G.~D.
  1997, \aj, 114, 1858

\bibitem[{{Rahman} {$et~al$.}(2007){Rahman}, {Howell}, {Helou}, {Mazzarella},
  \& {Buckalew}}]{Rahmanetal2007}
{Rahman}, N., {Howell}, J.~H., {Helou}, G., {Mazzarella}, J.~M., \& {Buckalew},
  B. 2007, \apj, 663, 908

\bibitem[{{Reshetnikov} {$et~al$.}(2010){Reshetnikov}, {Moiseev}, \&
  {Sotnikova}}]{Reshetnikovetal2010}
{Reshetnikov}, V.~P., {Moiseev}, A.~V., \& {Sotnikova}, N.~Y. 2010, \mnras,
  406, L90

\bibitem[{{Saha} \& {Elmegreen}(2016)}]{SahaElmegreen2016}
{Saha}, K., \& {Elmegreen}, B. 2016, \apjl, 826, L21

\bibitem[{{Saha} \& {Jog}(2014)}]{Sahajog2014}
{Saha}, K., \& {Jog}, C.~J. 2014, \mnras, 444, 352

\bibitem[{{Saha} {$et~al$.}(2010){Saha}, {Tseng}, \& {Taam}}]{Sahaetal2010}
{Saha}, K., {Tseng}, Y.-H., \& {Taam}, R.~E. 2010, \apj, 721, 1878

\bibitem[{{Schawinski} {$et~al$.}(2009){Schawinski}, {Lintott}, {Thomas},
  {Sarzi}, {Andreescu}, {Bamford}, {Kaviraj}, {Khochfar}, {Land}, {Murray},
  {Nichol}, {Raddick}, {Slosar}, {Szalay}, {VandenBerg}, \&
  {Yi}}]{Schawinski09}
{Schawinski}, K., {Lintott}, C., {Thomas}, D., {$et~al$.} 2009, \mnras, 396,
  818

\bibitem[{{Schlegel} {$et~al$.}(1998){Schlegel}, {Finkbeiner}, \&
  {Davis}}]{Schlegel98}
{Schlegel}, D.~J., {Finkbeiner}, D.~P., \& {Davis}, M. 1998, \apj, 500, 525

\bibitem[{{Sersic}(1968)}]{Sersic1968}
{Sersic}, J.~L. 1968, {Atlas de Galaxias Australes}

\bibitem[{{Tandon} {$et~al$.}(2017){Tandon}, {Subramaniam}, {Girish}, {Postma},
  {Sankarasubramanian}, {Sriram}, {Stalin}, {Mondal}, {Sahu}, {Joseph},
  {Hutchings}, {Ghosh}, {Barve}, {George}, {Kamath}, {Kathiravan}, {Kumar},
  {Lancelot}, {Leahy}, {Mahesh}, {Mohan}, {Nagabhushana}, {Pati}, {Kameswara
  Rao}, {Sreedhar}, \& {Sreekumar}}]{Tandonetal2017a}
{Tandon}, S.~N., {Subramaniam}, A., {Girish}, V., {$et~al$.} 2017, \aj, 154,
  128

\bibitem[{{Tandon} {$et~al$.}(2020){Tandon}, {Postma}, {Joseph}, {Devaraj},
  {Subramaniam}, {Barve}, {George}, {Ghosh}, {Girish}, {Hutchings}, {Kamath},
  {Kathiravan}, {Kumar}, {Lancelot}, {Leahy}, {Mahesh}, {Mohan},
  {Nagabhushana}, {Pati}, {Rao}, {Sankarasubramanian}, {Sriram}, \&
  {Stalin}}]{Tandonetal2020}
{Tandon}, S.~N., {Postma}, J., {Joseph}, P., {$et~al$.} 2020, \aj, 159, 158

\bibitem[{{Tody}(1993)}]{Tody93}
{Tody}, D. 1993, in Astronomical Society of the Pacific Conference Series,
  Vol.~52, Astronomical Data Analysis Software and Systems II, ed. R.~J.
  {Hanisch}, R.~J.~V. {Brissenden}, \& J.~{Barnes}, 173

\bibitem[{{Vaghmare} {$et~al$.}(2015){Vaghmare}, {Barway}, {Mathur}, \&
  {Kembhavi}}]{Vaghmareetal2015}
{Vaghmare}, K., {Barway}, S., {Mathur}, S., \& {Kembhavi}, A.~K. 2015, \mnras,
  450, 873

\bibitem[{{Vaghmare} {$et~al$.}(2018){Vaghmare}, {Barway}, {V{\"a}is{\"a}nen},
  {Ramphul}, {Wadadekar}, \& {Kembhavi}}]{Vaghmareetal2018}
{Vaghmare}, K., {Barway}, S., {V{\"a}is{\"a}nen}, P., {$et~al$.} 2018, \mnras,
  480, 4931

\bibitem[{{van den Bosch} {$et~al$.}(2008){van den Bosch}, {Aquino}, {Yang},
  {Mo}, {Pasquali}, {McIntosh}, {Weinmann}, \& {Kang}}]{vandenBoschetal2008}
{van den Bosch}, F.~C., {Aquino}, D., {Yang}, X., {$et~al$.} 2008, \mnras, 387,
  79

\bibitem[{{van Driel} \& {van Woerden}(1991)}]{vanDriel1991}
{van Driel}, W., \& {van Woerden}, H. 1991, \aap, 243, 71

\bibitem[{{van Eymeren} {$et~al$.}(2011){van Eymeren}, {J{\"u}tte}, {Jog},
  {Stein}, \& {Dettmar}}]{vanEymerenetal2011}
{van Eymeren}, J., {J{\"u}tte}, E., {Jog}, C.~J., {Stein}, Y., \& {Dettmar},
  R.~J. 2011, \aap, 530, A30

\bibitem[{{Yozin} \& {Bekki}(2014)}]{YozinBekki2014}
{Yozin}, C., \& {Bekki}, K. 2014, \mnras, 439, 1948

\end{thebibliography}

%


\end{document}